\documentclass[12pt, letterpaper]{article}

\pdfoutput=1

\usepackage{amsmath}
\usepackage{amsfonts}
\usepackage{amssymb}
\usepackage{comment}
\usepackage[colorlinks,citecolor=blue,urlcolor=blue,bookmarks=false,hypertexnames=true]{hyperref}
\usepackage{cite}

\usepackage{graphicx, physics, subcaption}
\usepackage{verbatim }

\numberwithin{equation}{section}

\usepackage{color}

\usepackage{amsmath, amssymb, graphics, epsfig, graphicx}
\usepackage{epsf}
\usepackage{epstopdf}
\usepackage {amssymb}
\newcommand{\nc}{\newcommand}

\nc{\cN}{ {\cal{N}} }

\nc{\ba}{\begin{eqnarray}}
\nc{\ea}{\end{eqnarray}}

\begin{document}

\vspace{5mm}
\vspace{0.5cm}
\begin{center}

\def\thefootnote{\fnsymbol{footnote}}

{\bf\large {{Induced Gravitational Waves from Non-attractor Inflation\\ and
NANOGrav data } }}
\\[0.5cm]

{Amin Nassiri-Rad \footnote{amin.nassiriraad@ipm.ir}, 
Kosar Asadi \footnote{k.asadi@ipm.ir}
}
\\[.7cm]

{\small \textit{School of Astronomy, 
		Institute for Research in Fundamental Sciences (IPM) \\ 
		P.~O.~Box 19395-5531, Tehran, Iran }} \\

\end{center}

\vspace{.8cm}

\hrule \vspace{0.3cm}


\begin{abstract}
In this paper, we investigate the scalar-induced gravitational waves in single-field non-attractor inflation for the Pulsar Timing Arrays data. Our model comprises three phases of inflation: the first and third phases are slow-roll inflation, while the second phase is a period of non-attractor inflation. We analyze the model's predictions for various values of the sound speed $c_s$ and examine the  sharp transitions to the final attractor phase. Furthermore, we study the model's predictions for NANOGrav observations and future gravitational wave observations. We also calculate the non-Gaussianity parameter $f_{NL}$ for the non-attractor setup with a general sound speed and the sharpness parameter.

\end{abstract}
\vspace{0.5cm} \hrule
\def\thefootnote{\arabic{footnote}}
\setcounter{footnote}{0}
\newpage

\section{Introduction}

The detection of gravitational waves (GWs) has indeed revolutionized our understanding of physics on large scales and in non-linear regimes. Recent advancements have led to significant progress in detecting stochastic GWs in the range of 10 nHz, as observed by NANOGrav \cite{NANOGrav:2023gor} and other experiments \cite{Reardon:2023gzh,Antoniadis:2023ott,Xu:2023wog}. The NANOGrav data provides valuable insights into the astrophysical origins of GWs, including the merging of binary black holes. Another intriguing possibility is the existence of induced gravitational waves, which are sourced by scalar perturbations on small scales during inflation \cite{Antoniadis:2023zhi,NANOGrav:2023hvm,Vilenkin:1981zs,Caldwell:1991jj,Vilenkin:1981bx,Kibble:1976sj,Hindmarsh:2013xza,Caprini:2007xq,Kamionkowski:1993fg,Kosowsky:1992rz}.

To explain the NANOGrav data via cosmological sources, one needs to amplify the scalar perturbations during inflation \cite{Ananda:2006af,Baumann:2007zm,Bugaev:2009zh,Assadullahi:2009nf,Alabidi:2012ex,Cai:2018dig,Pi:2020otn,Balaji:2022dbi,Talebian:2022cwk,Domenech:2021ztg}. The amplification of scalar perturbations is also a mechanism for the production of primordial black holes (PBHs), which may explain dark matter as well\cite{Carr:2016drx,Carr:2020xqk,Sasaki:2018dmp,Ozsoy:2023ryl,Byrnes:2021jka}. The GW perturbations are then sourced by the square of the scalar perturbations. If the amplitude of the scalar perturbations at small scales is much higher than the corresponding amplitude at CMB scales, then there is a chance that the induced GWs are large enough to explain NANOGrav data.

The Ultra slow-roll (USR) inflation model is a well-known approach to amplify scalar perturbations, which can lead to the formation of PBHs \cite{Ivanov:1994pa,Garcia-Bellido:2017mdw,Biagetti:2018pjj,Ragavendra:2020sop,Di:2017ndc,Liu:2020oqe,Hooshangi:2022lao,Hooshangi:2023kss,Ghoshal:2023wri}. In this model, the potential is constant, resulting in an exponential decrease in the velocity of the inflaton over time. Consequently, this leads to an increase in the curvature perturbation \cite{Kinney:2005vj,Morse:2018kda,Lin:2019fcz,Namjoo:2012aa}.
It is important to note that the USR model violates the Maldacena consistency condition \cite{Maldacena:2002vr,Creminelli:2004yq}. Several studies have explored this violation, including \cite{Martin:2012pe,Chen:2013aj,Chen:2013eea,Akhshik:2015rwa,Mooij:2015yka,Bravo:2017wyw,Finelli:2017fml,Pi:2022ysn}. According to \cite{Namjoo:2012aa}, the non-Gaussianity parameter in this model is $f_{NL}=\frac{5}{2}$. In \cite{Cai:2018dkf} this result is extended by introducing a sharpness parameter that characterizes the final attractor phase.

In an alternative scenario, the perturbations can experience growth during the non-attractor phase, even with an arbitrary sound speed  $c_s$ \cite{Chen:2013aj,Chen:2013eea}. In this study, we extend the approach proposed in \cite{Firouzjahi:2023lzg} by considering the arbitrary sound speed $c_s$ as an additional degree of freedom. Our objective is to investigate how this model can account for the NANOGrav data. In our model, the Lagrangian of the inflaton is an arbitrary function of the kinetic term $X$, denoted as $P(X)$, with a constant potential. The sound speed can have non-trivial effects on the generalization of the results obtained in \cite{Firouzjahi:2023lzg}. Specifically, our model comprises three inflationary phases: two slow-roll phases at the beginning and end, and a non-attractor phase in between. During the slow-roll phases, the sound speed is set equal to unity. In this work, we also explore non-Gaussianity and demonstrate that it depends on both the sharpness parameter $h$ and the number of e-folds associated with the non-attractor phase. Notably, when $c_s=1$, our results reduce to the standard result obtained in \cite{Cai:2018dkf}. Furthermore, if the transition to the final attractor phase is very mild, non-Gaussianity approaches zero.
 
The researches on ultra slow-roll inflation has recently focused on the loop effects of this model on the CMB scale \cite{yokoyama,Kristiano:2023scm,Riotto:2023hoz,Riotto:2023gpm,Choudhury:2023jlt,Choudhury:2023rks,Firouzjahi:2023aum,hr,Motohashi:2023syh,f,Tasinato:2023ukp,Firouzjahi:2023btw,Fumagalli:2023hpa,Tada:2023rgp}. The main message conveyed by these works is that loop corrections can become significant and affect long CMB scales perturbations, depending on the duration of ultra slow-roll inflation. In particular, it is claimed in \cite{Firouzjahi:2023aum} that if the transition to the final attractor phase is mild, then loop corrections will become harmless. However, other criticisms have been raised that loop corrections are absent \cite{Riotto:2023hoz,Riotto:2023gpm,Fumagalli:2023hpa}. This question remains open, and future studies will be needed to investigate the loop effects in our model.

The paper is structured as follows. In Section \ref{modelsec}, we present our model and generalize the results of  \cite{Firouzjahi:2023lzg}  to the case with a general sound speed. We calculate the non-Gaussianity for this model and study its consistency in confrontation with the observations of PBHs. In Section \ref{PTAobserv}, we compare our results with the NANOGrav data and discuss the differences compared to the USR model. Our main focus is to investigate the predictions of the model for various values of $c_s$ and see which ones fit better to the data qualitatively. We conclude in Section \ref{conclusion}.

\vspace{0.5cm}

\section{The Model}
\label{modelsec}

In this section, we introduce our model for the scalar-induced gravitational waves as the source of pulsar timing arrays (PTAs) observations. Our setup comprises three stages: two slow-roll inflation stages at the beginning and end, and a non-attractor phase in the middle. This model is a generalization of the work presented in  \cite{Firouzjahi:2023lzg}, where the middle phase allows for an arbitrary sound speed, $c_s$. Specifically, the Lagrangian in the middle phase is given by
\begin{equation}
    \mathcal{L}_2=P(X)-V_0,
\end{equation}
where $V_0$ is a constant, and $X=\frac{\dot\phi^2}{2}$ is the kinetic energy. The sound speed in terms of $P(X)$ is given by \cite{Chen:2006nt}
\begin{equation}
    c_s^2\equiv\frac{P_{,X}}{P_{,X}+2XP_{,XX}}.
\end{equation}
As $P(X)=X$ in the first and third stages, the sound speed is set to unity ($c_s=1$) in these phases. In the non-attractor phase, the evolution of $\phi$ at the background level is given by \cite{Chen:2013eea}
\begin{equation}
\Ddot\phi+3H\dot\phi c_s^2=0.
\end{equation}

In the first slow-roll phase, the mode function in terms of conformal time $\tau$  is given by 
\begin{equation}
\mathcal{R}_1= \frac{H e^{-i k \tau} (1+i k\tau)}{2 \sqrt{k^3 \text{$\epsilon_i $}}}.
\end{equation}
Here we assume that $\epsilon_i$ represents the slow roll parameter at the first phase, which is assumed to be nearly constant. We also consider the Bunch-Davies initial condition for the primary mode function. Throughout this paper, we assume $M_P=1$ without loss of generality. In the middle stage, as the sound speed is arbitrary, the mode function will be more complicated. Following \cite{Akhshik:2015nfa}, the mode function in the non-attractor phase evolves as
\begin{equation}
\label{modeevolv}
  v_k''+(c_s^2 k^2-\frac{(\eta-2 s+3)^2-1}{4\tau^2})v_k=0,  
\end{equation}
where we define the variable
$v_k=z \mathcal{R}_k$ , $z=\frac{2\epsilon a^2} {c_s^2}$  and $s=\frac{\dot{c_s}}{H c_s}$. Here, $a$ represents the scale factor. Additionally, we have the first and second slow-roll parameters, $\epsilon$ and $\eta$, respectively. They are defined as $\epsilon=-\frac{\dot{H}}{H^2}$ and $\eta = \frac{\dot{\epsilon}}{H \epsilon}$ (a 'dot' represents the differentiation with respect to the cosmic time).
In the non-attractor phase, we assume that the phase begins at $\tau=\tau_i$ and ends at $\tau=\tau_e$. Based on this assumption, it can be shown that the first and second slow-roll parameters during this phase with a constant potential are expressed as
\begin{equation}
    \epsilon_n(\tau)=\text{$\epsilon_i $} \left(\frac{\tau}{\tau_i}\right)^{-\eta },
\end{equation}
\begin{equation}
    \eta=-3(1+c_s^2).
\end{equation}
The general solution to \eqref{modeevolv}  in the non-attractor phase is given as
\begin{equation}
   \mathcal{R}_2= \frac{iH c_s \sqrt{-\tau ^3}}{2\sqrt{\epsilon}_i}\left(\frac{\tau _i}{\tau }\right)^{-\frac{\eta }{2}}  \bigg[\alpha _k H_v^{(1)}(-c_s k \tau )+\beta _k H_v^{(2)}(-c_s k \tau )\bigg], 
\end{equation}
where $\nu=\frac{-3 c_s^2}{2}$. $\alpha_k$ and $\beta_k$ are constants that should be determined by the matching condition at $\tau=\tau_i$. As the sound speed is different in the middle stage 
the matching conditions are as follows\cite{Namjoo:2012xs}
\begin{equation}
\label{matching}
    \begin{split}
&R_k(\tau_i^-) =R_k(\tau_i^+)\\
&R'_k(\tau_i^-) =\frac{R'_k(\tau_i^+)}{c_s^2}.
    \end{split}
\end{equation}
  we can determine the values of $\alpha_k$ and $\beta_k$ as
\begin{equation}
 \alpha_k= \frac{i e^{-i k \tau _i} }{2 \sqrt{-k^3 \tau _i^3}}\bigg[k \tau _i \left(k \tau _i-i\right) H_{1+\nu}^{(2)}\left(-k c_s \tau _i\right)+i c_s \left(3+k \tau _i \left(k \tau _i+3 i\right)\right) H_{\nu}^{(2)}\left(-k c_s \tau _i\right)\bigg], 
\end{equation}
\begin{equation}
 \beta_k=-\frac{i e^{-i k \tau _i} }{2 \sqrt{-k^3 \tau _i^3}}\bigg[k \tau _i \left(k \tau _i-i\right) H_{1+\nu}^{(1)}\left(-c_s k \tau _i\right)+i c_s \left(3+k \tau _i \left(k \tau _i+3 i\right)\right) H_{\nu}^{(1)}\left(- c_s k \tau _i\right)\bigg].
\end{equation}
One can determine the mode function at the third stage by utilizing $\alpha_k$ and $\beta_k$. It is possible to verify that when $c_s=1$, the aforementioned expressions reduce to the findings of \cite{Firouzjahi:2023aum}. The first slow-roll parameter in the final stage is given by \cite{Cai:2018dkf} 
\begin{equation}
 \epsilon(\tau)= \epsilon _e \left[\frac{h}{6}-\left(\frac{h}{6}+1\right) \frac{\tau ^3}{\tau_e^3}\right]^2.
\end{equation}
Here, the parameter $h$ determines the sharpness of the transition to the final attractor phase. A higher value of $|h|$ leads to a quicker transition. In our model, $h$ is defined as\cite{Cai:2018dkf,Firouzjahi:2023aum}
\begin{equation}
  h=-6\sqrt{\frac{\epsilon_V}{\epsilon_e}}, 
\end{equation}
where $\epsilon_e$ represents the slow-roll parameter at the end of the non-attractor phase, and $\epsilon_V=\frac{1}{2}(\frac{V'(\phi_e)}{V(\phi_e)})^2$. 

One can use these results to construct the mode function. The general mode function in the final attractor phase is expressed as
\begin{equation}
    \mathcal{R}_3=\frac{H}{2 \sqrt{k^3 \epsilon (\tau )}} \bigg[c_k e^{-i k \tau } (1+i k \tau )+d_k e^{i k \tau } (1-i k \tau )\bigg]
\end{equation}
The constants $c_k$ and $d_k$ are determined by matching conditions similar to \eqref{matching} at $\tau=\tau_e$. Due to their complexity, we do not provide explicit expressions for them. The power spectrum at the end of inflation can be obtained using the following relation
\begin{equation}
    \mathcal{P}_\mathcal{R}=\frac{k^3}{2\pi^2} |\mathcal{R}_3|^2\Big|_{\tau=0}
\end{equation}

\begin{figure}
    \begin{subfigure}[t]{0.48\textwidth}
        \includegraphics[width=\linewidth]{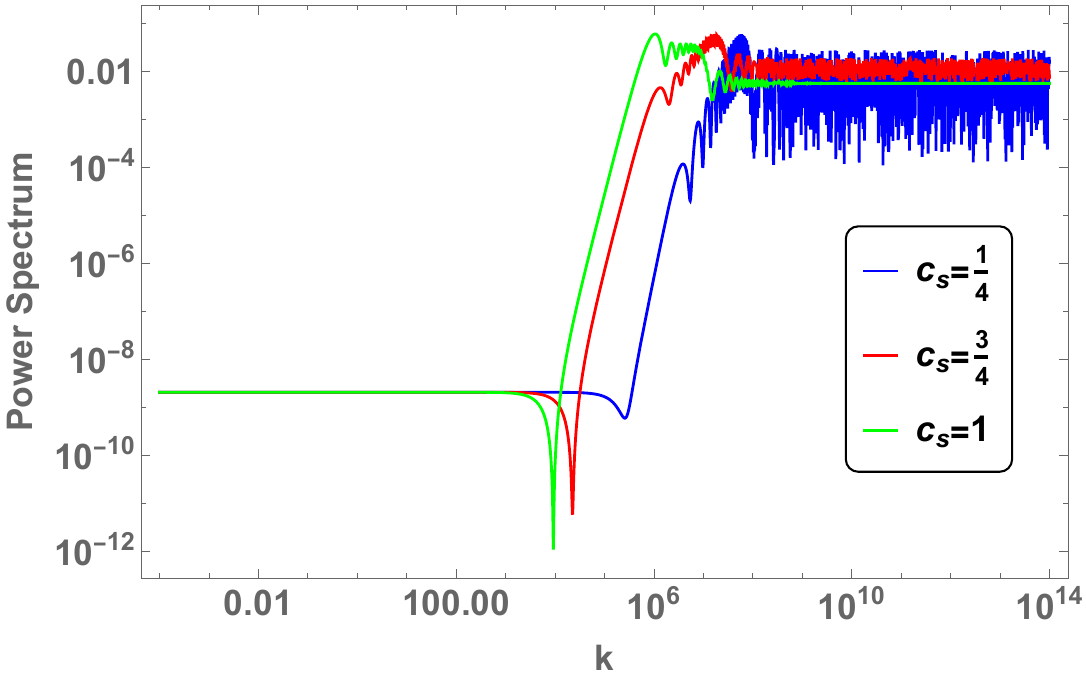} 
    \end{subfigure}
    \hfill
    \begin{subfigure}[t]{0.48\textwidth}
        \includegraphics[width=\linewidth]{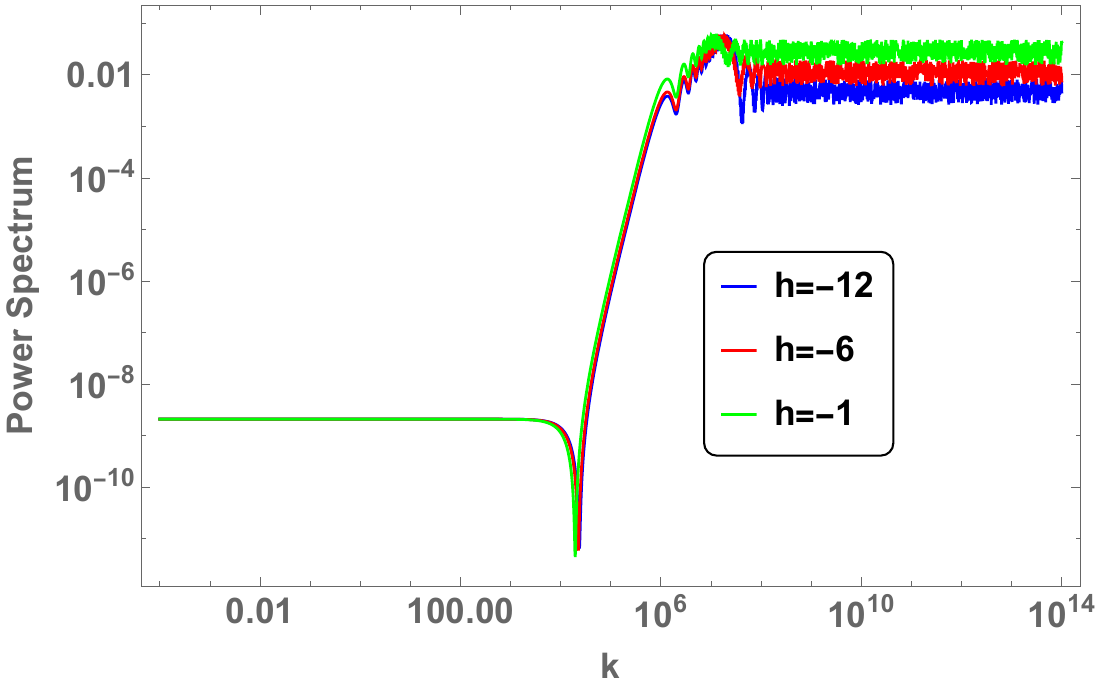} 
    \end{subfigure}
    \caption{
    Left: The power spectrum for different values of $c_s$ and $h=-6$. Other parameters are held constant to ensure that the peak for different sound speeds does not grow significantly.
    Right: The power spectrum for $c_s=\frac{3}{4}$ and different values of $h$. } 
    \label{power}
\end{figure}

\begin{figure} 
    \centering
    \includegraphics[scale=0.4]{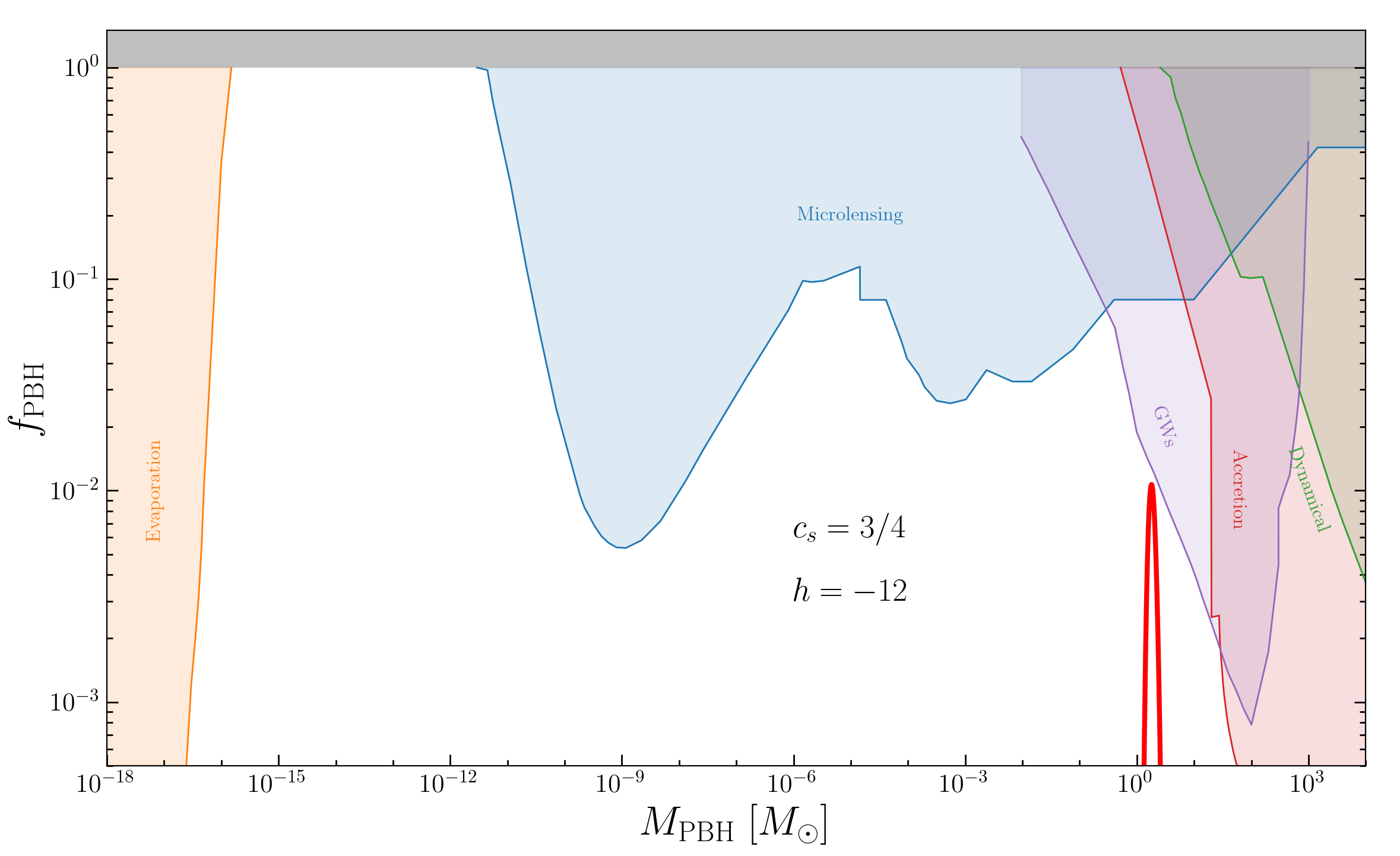}
    \caption{This figure illustrates the behavior of the PBH mass fraction ($f_{\text{PBH}}$) for $c_s = \frac{3}{4}$ and $h = -12$. It is worth noting that this behavior is similar for other values of $c_s $ and $h$, so we have consolidated them into a single figure.}
    \label{fPB}
\end{figure}

The two graphs in Fig. \ref{power} illustrate the power spectrum for various combinations of $c_s$ and $h$. To calculate the power spectrum, we select $\tau_i$  such that the non-attractor phase commences about 16 e-folds after the CMB mode exits the horizon. Additionally, we employ COBE normalization to establish the power spectrum on CMB scales. In simpler terms, by satisfying the matching condition, it can be demonstrated that on super-horizon scales, we observe
 \begin{equation}
 \label{COBE}
     \lim_{k\rightarrow 0} \frac{k^3}{2\pi^2} |\mathcal{R}_3|^2=\frac{H^2}{8\pi^2\epsilon_i}
 \end{equation}
The values of $H$ and $\epsilon_i$ are carefully selected such that equation \eqref{COBE} yields a power spectrum $\mathcal{P}_\mathcal{R}\simeq 2\times 10^{-9}$ for $k_{CMB}=0.05 \text{Mpc}^{-1}$.
As depicted in Fig. \ref{power}, the power spectrum exhibits similar characteristics to those described in \cite{Firouzjahi:2023lzg}, which will be elaborated upon in detail. Let $\Delta N$  denote the e-fold duration of the non-attractor phase. For $|\tau_e|\ll |\tau_i|$ and $k |\tau_e|\ll 1$ , it can be demonstrated that
\begin{equation}
\label{xlarge}
\begin{split}
\mathcal{P}_\mathcal{R}=&\mathcal{P}_\mathcal{\text{CMB}}\frac{(6-h)^2}{h^2 }e^{6 c_s^2 \Delta N}\times\\&\Bigg|(1+i x) \, _0F_1\left(;\frac{3 c_s^2}{2};-\frac{1}{4} x^2 c_s^2\right)-\frac{1}{3} (3+x (x+3 i)) \, _0F_1\left(;\frac{3 c_s^2}{2}+1;-\frac{1}{4} x^2 c_s^2\right)\Bigg|^2,
    \end{split}
\end{equation}
where $x=-k \tau_i$ and $_0F_1$ is the hypergeometric function. From the above equation, we observe that the exponential growth of the power spectrum is modified to $\exp(6 c_s^2 \Delta N)$. This indicates that the peak of the non-attractor phase is shorter compared to the case where $c_s=1$. Additionally, the power spectrum is proportional to $\frac{(6-h)^2}{h^2 }$. 
It is worth noting that the scaling behavior of the power spectrum in the attractor phase is identical to that observed in ultra slow-roll inflation.
Another interesting feature to study is the behavior of the power spectrum in the limit $|k\tau_i|\ll1$, which pertains to modes that exited the horizon at early times. Given the power spectrum at the end of inflation, one can observe that the power spectrum at CMB scales is modified by
\begin{equation}
\label{powerx2}
    \mathcal{P}_{\mathcal{R}}(k,\tau=0)\simeq \mathcal{P}_{\text{CMB}} \bigg[1+\frac{2 x^2 \left( c_s^2+1\right) \left(6 -h\right)e^{3c_s^2\Delta N} }{ h  \left(3 c_s^2+2\right)} \bigg].
\end{equation}
 As we see in Fig. \ref{power} in the dip, the power spectrum decreases considerably. To estimate the position of the dip, we expect Eq. \eqref{powerx2} goes to zero at dip and one has
 \eqref{powerx2}  
\begin{equation}
    k|\tau_i|\sim e^{-3c_s^2\Delta N/2}\sqrt{\frac{h \left(3 c_s^2+2\right)}{2\left( c_s^2+1\right) \left(h - 6\right)}}.
\end{equation}
The above expression also demonstrates that the position of the dip shifts to the right as the absolute value of $|h|$ increases.
One can also check the behaviour of power spectrum for the modes very deep inside horizon. For $1\ll k|\tau_e|$ one can check that 

\begin{equation}
\begin{split} 
    \mathcal{P}_\mathcal{R}\simeq & \frac{9 H^2 e^{3\Delta N (1+c_s^2)}}{8 \pi ^2 h^2 c_s^2 \epsilon _i} \times\\& 
    \Bigg|\left(c_s-1\right) e^{2 i k c_s (\tau _i-\tau_e)} \left(\cos \left(k \tau _e\right)+i c_s \sin \left(k \tau _e\right)\right)+\left(c_s+1\right)  \left(\cos \left(k \tau _e\right)-i c_s \sin \left(k \tau _e\right)\right)\Bigg|^2 \,.
\end{split}
\end{equation}
The given expression oscillates for large values of k when $c_s\neq 1$. However, if we consider $c_s=1$, the power spectrum remains constant for $1\ll k|\tau_e|$, as depicted in the left panel of Figure \ref{power}. In particular, the expression reduces to the following one by setting $c_s=1$
\begin{equation}
    \mathcal{P}_\mathcal{R}\simeq \frac{9 H^2 e^{6 \Delta N}}{2 \pi ^2 h^2  \epsilon _i}\,.
\end{equation}
To fit the parameters of the model with PTAs data, we also consider the observed values for PBH formation, as shown in Fig. \ref{fPB}. To begin, we focus on the mass fraction of PBHs, which is given by\cite{Garcia-Bellido:2016dkw}
\begin{equation}
\label{betaestimate}
    \beta\simeq \frac{1}{2}\text{Erfc}\bigg(\frac{\mathcal{R}_c}{\sqrt{2\mathcal{P}_\mathcal{R}}}\bigg).
\end{equation}
Here, $\mathcal{R}_c$ is an $O(1)$ parameter that denotes the critical value of the curvature perturbation for PBH formation. In this model, we have chosen $\mathcal{R}_c=1.59$. Additionally, the ratio of PBHs to the dark matter density, $f_{PBH}$, is expressed as \cite{Sasaki:2018dmp}  
\begin{equation}
    f_{\text{PBH}}(M_{\text{PBH}})\simeq 2.7\times10^8 \bigg(\frac{M_{\text{PBH}}}{M_\odot}\bigg)^{-\frac{1}{2}}\beta.
\end{equation}
To determine $\frac{M_{\text{PBH}}}{M_\odot}$ in terms of the number of e-folds, one can use the following relation
\cite{Ozsoy:2023ryl}
\begin{equation}
  \frac{M_{\text{PBH}}}{M_\odot} =7.7 \times 10^{17} \exp (-2(N_{\text{max}}-N_{\text{CMB}})).
\end{equation}
In our model, $N_{\text{max}}$ represents the number of e-folds at which the power spectrum reaches its maximum and $N_{CMB}$ is the number of e-folds at which CMB mode exits the horizon. For each value of $c_s$ and $h$, our model has three free parameters:  $h$, $\tau_i$ and $\Delta N$. We have carefully selected these parameters such that for each $c_s$  the frequency of the peak in the power spectrum is comparable to the frequency of PTAs data. In Fig. \ref{fPB}  we have presented constraints on $f_{\text{PBH}}$ for different values of $c_s$. As the constraints on $f_{\text{PBH}}$ for different parameters were somewhat similar, we have consolidated them into a single figure. When $h=-1$, the amplitude of oscillations does not decay for large momentum values, as observed in Figure \ref{power}. Since the approximation of $\beta$ in Equation \eqref{betaestimate} may not be valid in this limit, we do not study the effect of mild transitions.
Before ending this section, we delve into the study of the bispectrum for non-attractor models with arbitrary sound speed $c_s$ in the next subsection.

\subsection{Bispectrum}

In this subsection, we evaluate the bispectrum for the modes that exit the horizon during the intermediate non-attractor phase. To calculate the bispectrum, one can adopt a two-phase inflationary framework, where the first phase corresponds to the non-attractor phase, and the second phase corresponds to the slow-roll phase \cite{Firouzjahi:2023aum}. Since the mode functions during non-attractor inflation do not exhibit scale invariance, we can consider the following option for the mode function
\begin{equation}
 \mathcal{R}_1(\tau)= \frac{ H c_s \sqrt{-\pi \tau ^3} }{2 \sqrt{2\epsilon _n(\tau)}}\bigg[c_{+} H_{\nu }^{(1)}\left(- c_s k \tau  c_s\right)+c_{-} H_{\nu }^{(1)}\left(- c_s k \tau \right)\bigg].
\end{equation}
The mode function in the second phase can be obtained using the following expressions for the coefficients
\begin{equation}
\begin{split}
&c_{\pm}=\frac{1}{2} \left(\frac{k}{\lambda}\right)^{-\frac{3}{2} \left(c_s^2-1\right)} \left[\left(\frac{k}{\lambda}\right)^{3 (c_s^2-1)} \pm 1\right]\,,
\end{split}
\end{equation}
where $\lambda$ is an arbitrary conststant. It is straightforward to verify that with these choices of coefficients, the mode function satisfies the normalization condition
\begin{equation}
    c_{+}^2-c_{-}^2=1.
\end{equation}
The mode function in the second phase is then given by
\begin{equation}
   \mathcal{R}_2(\tau)=\frac{H}{2\sqrt{k^3 \epsilon (\tau )}}\bigg[a_k e^{-i k \tau } (1+i k \tau )+b_k e^{i k \tau } (1-i k \tau )\bigg], 
\end{equation}
with $a_k$ and $b_k$ to be determined by the matching conditions. Based on the definition of $f_{NL}$, we have the following relation
\begin{equation}
\left<\mathcal{R}_{k_1}\mathcal{R}_{k_2}\mathcal{R}_{k_3}\right>= (2\pi)^3 \delta^3 \bigg(\sum_{i} \overrightarrow{k_i}\bigg) \frac{6}{5}f_{NL}\big(P_{k_1}P_{k_2}+P_{k_2}P_{k_3}+P_{k_3}P_{k_1}\big),
\end{equation}
where $P_k=|R_k(\tau=0)|^2$.
In order to calculate the bispectrum, we require the Hamiltonian up to third order. Additionally, it is more convenient to compute $f_{\text{NL}}$ using the effective field theory approach. It can be shown that  The third-order Hamiltonian, which neglects sub-leading gradient interactions, can be expressed as follows \cite{Akhshik:2015nfa}
\begin{equation}
\begin{split}
  &H_3^{(1)}=  -\frac{a^2 \eta  H^3 \epsilon  \pi'^2\pi}{c_s(\tau)^2}\,,\\
 &H_3^{(2)}= \left[1-\frac{1}{c_s(\tau)^2}\right]a H^2 \epsilon \pi'^3\,,\\
&H_3^{(3)}=2H^2a\epsilon\frac{c_s'(\tau)}{ c_s^3(\tau)}\pi'^2\pi,
  \end{split}
\end{equation}
where $c_s(\tau)$ is the time dependent sound speed. The total third order Hamiltonian is expressed as $H_3=H_3^{(1)}+H_3^{(2)}+H_3^{(3)}$. Here, the prime denotes the derivative with respect to conformal time, and $\pi$ represents the Goldstone boson, which is related to the curvature perturbation according to the relations established in \cite{Maldacena:2002vr,Cheung:2007sv}
\begin{equation}
    \mathcal{R}_k=-H \pi_k+H\pi_k\dot\pi_k+\frac{\dot H}{2}\pi_k^2 \, .
\end{equation}
Since our focus is on the three-point correlation function at the end of inflation, the higher-order terms are suppressed within the slow-roll approximation. Therefore, we can utilize the aforementioned relation at the linear order, leading to an approximate expression for the three-point correlation function as follows
\begin{equation}
    \left<\mathcal{R}_{k_1}\mathcal{R}_{k_2}\mathcal{R}_{k_3}\right>\simeq-H^3\left<\pi_{k_1}\pi_{k_2}\pi_{k_3}\right>.
\end{equation}
By employing the in-in formalism, we can express the three-point correlation function as follows

\begin{equation}
\label{inin}
    \left<\pi_{k_1}(0)\pi_{k_2}(0)\pi_{k_3}(0)\right>=-2 \Im \int_{\tau_i}^0 d\tau \left<H_3(\tau)\pi_{k_1}(0)\pi_{k_2}(0)\pi_{k_3}(0)\right>.
\end{equation}
The integral mentioned above can be divided into three parts: $\tau_i<\tau<\tau_e$, $\tau=\tau_e$  and$\tau_e<\tau<0$. At $\tau=\tau_e$ the Hamiltonian at the transition time contributes to the three-point function.
To compute the contribution of each part, we utilize the mode functions corresponding to the non-attractor phase in the first part, and those associated with the slow-roll phase in the second part. For $\tau_i<\tau<\tau_e$ we have the non-attractor phase and $c_s(\tau)=c_s$. In this part both $H_3^{(1)}$ and $H_3^{(2)}$ contribute to the three-point function and we have

\begin{equation}
\frac{\left<\mathcal{R}_{k_1}\mathcal{R}_{k_2}\mathcal{R}_{k_3}\right>^{(1)}_{\text{non-att}} }{2\pi^2\mathcal{P}_\mathcal{R} }=(2\pi)^3 \delta^3 \bigg(\sum_{i} \overrightarrow{k_i}\bigg)\frac{-3h \left(c_s^2+1\right) (h+12)}{2 (h-6)^2}\left(\frac{1}{k_1^3}\frac{1}{k_2^3}+ 2 \text{ perms}\right),
\end{equation}

\begin{equation}
\frac{\left<\mathcal{R}_{k_1}\mathcal{R}_{k_2}\mathcal{R}_{k_3}\right>^{(2)}_{\text{non-att}} }{2\pi^2\mathcal{P}_\mathcal{R} }=(2\pi)^3 \delta^3 \bigg(\sum_{i} \overrightarrow{k_i}\bigg)\frac{9 h^2 c_s^2 \left(c_s^2-1\right)}{4 (h-6)^2}\left(\frac{1}{k_1^3}\frac{1}{k_2^3}+2 \text{ perms}\right),
\end{equation}
with $\mathcal{P}_\mathcal{R}= \lim_{k\rightarrow 0} \frac{k^3}{2\pi^2} |\mathcal{R}_3|^2$. To clarify, we use the term ‘non-att’ to refer to the non-attractor phase. On the other hand for $\tau_e<\tau<0$ we have the slow roll phase. In this stage only $H_3^{(1)}$ contributes to the three-point function with $c_s(\tau)=1$ . Computing the three-point function in this part yields
\begin{equation}
\frac{\left<\mathcal{R}_{k_1}\mathcal{R}_{k_2}\mathcal{R}_{k_3}\right>^{(1)}_{\text{SR}}}{2\pi^2\mathcal{P}_\mathcal{R}}= (2\pi)^3 \delta^3 \bigg(\sum_{i} \overrightarrow{k_i}\bigg)\frac{6 h (h+6) e^{3 \Delta N \left(c_s^2-1\right)}}{(h-6)^2}\left(\frac{1}{k_1^3}\frac{1}{k_2^3}+2 \text{ perms}\right).
\end{equation}
Finally, we calculate the contribution of the transition term. Since the sound speed changes during the transition, we can express $c_s(\tau)^2$ as
\begin{equation}
  c_s(\tau)^2=c_s^2\theta(\tau_e-\tau)+\theta(\tau-\tau_e),
\end{equation}
with $\theta(x)$ being the step function. Writing $H_3^{(3)}=H^2a\epsilon\frac{(c_s(\tau)^2)'}{ c_s^4(\tau)}\pi'^2\pi$ and using \eqref{matching} at $\tau=\tau_e$, the above equation gives the contribution of transition term to the three-point function as

\begin{equation}
\frac{\left<\mathcal{R}_{k_1}\mathcal{R}_{k_2}\mathcal{R}_{k_3}\right>^{(3)}_{\text{transition}}}{2\pi^2\mathcal{P}_\mathcal{R}}= (2\pi)^3 \delta^3 \bigg(\sum_{i} \overrightarrow{k_i}\bigg)\frac{3 h (2 h+6) }{(h-6)^2} e^{-3 \Delta N(1-c_s^2)}(c_s^2-1)\left(\frac{1}{k_1^3}\frac{1}{k_2^3}+2 \text{ perms}\right).
\end{equation}

Taking into account all the above contributions the non-Gaussianity then reads as
\begin{equation}
\label{fnl}
f_{NL}=\frac{5h}{2(h-6)^2} \left[2 \left((h+3) c_s^2+3\right) e^{-3 \Delta N\left(1-c_s^2\right)}+\frac{ \left(-3 h c_s^4+(h-24) c_s^2-2 (h+12)\right)}{4 }\right]\,.
\end{equation}

The expression mentioned above violates the Maldacena consistency condition, which is similar to what happens in ultra slow-roll inflation. It is noteworthy that when $c_s=1$, the expression reduces to the result obtained in \cite{Firouzjahi:2023aum}. Furthermore, we can analyze the sharp transition limit where $h\rightarrow -\infty$. In this limit, the expression simplifies to
\begin{equation}
    f_{NL}=\frac{5}{8} \left[c_s^2 \left(8 e^{-3 \Delta N \left(1-c_s^2\right)}+1\right)-3 c_s^4-2\right]\,.
\end{equation}
It should be noted that the aforementioned outcome should not be mistaken for the outcome obtained in \cite{Chen:2013eea}. In that study, the non-attractor phase is assumed to have $\eta=-6$, which results in a non-constant potential. In contrast, in the present analysis, we observe that for $|h|\ll 1$, $f_{NL}$ approaches zero, which is similar to the result obtained in the case of ultra slow-roll inflation \cite{Firouzjahi:2023aum}.

According to \cite{Young:2013oia}, the Gaussian bound for the PBH constraints is applicable when $|f_{NL}|\mathcal{P}_\mathcal{R}\ll1$. For instance, with $h=-6$, $c_s=\frac{1}{4}$, and $\Delta N=3$, we obtain $f_{NL}\simeq 0.3$, and the Gaussian approximation constraint is satisfied with $\mathcal{P}_\mathcal{R}=\mathcal{O}(0.01)$. It is worth noting that the non-Gaussianity parameter presented in \eqref{fnl} corresponds to a model with two phases. However, if one were to inquire about the non-Gaussianity parameter in a model with three phases, our analysis for this case confirms that $f_{NL}=\mathcal{O}(\epsilon)$, which is negligible. Therefore, the aforementioned criterion is satisfied once again.

\section{PTAs observations and the scalar induced gravitational waves}
\label{PTAobserv}
\begin{figure} 
    \includegraphics[scale=0.85]{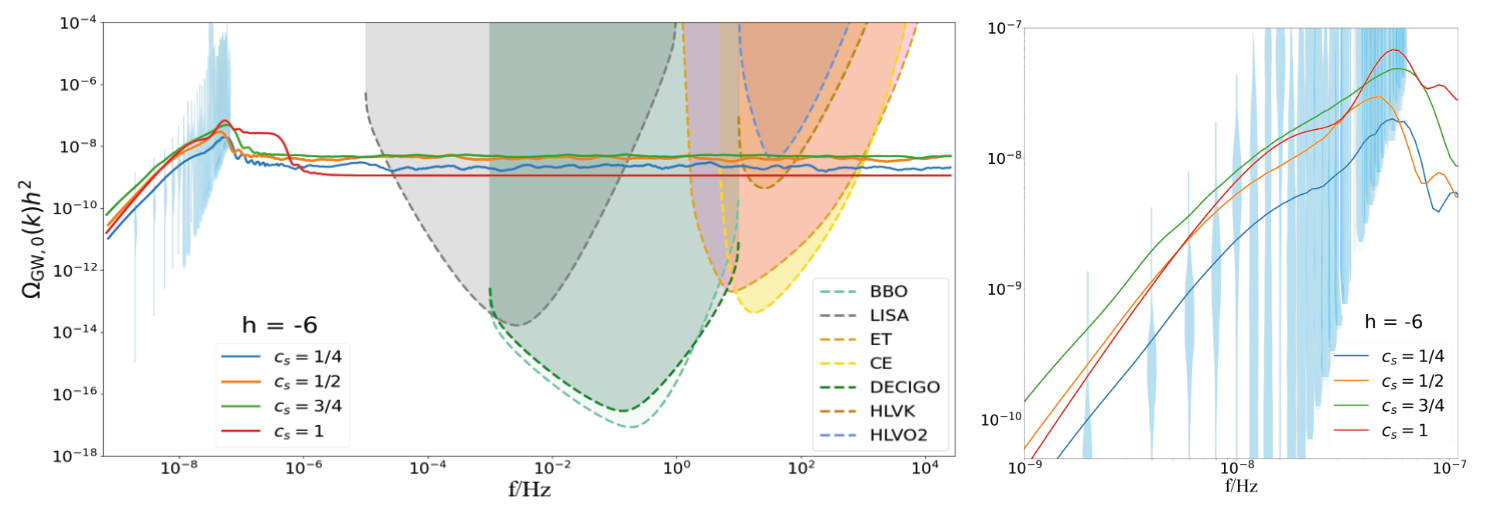}
    \caption{Comparison of the model's results with NANOGrav and future data for $h=-6$. It can be seen that $c_s=1/4$ has a better fit, particularly with the first violins, compared to other values of sound speed. Moreover, $c_s=1/2$ and $c_s=3/4$ lie in a more wider range of future predictions.}
    \label{Omega-obs-cs-h6}
\end{figure}

\begin{figure} 
    \includegraphics[scale=0.85]{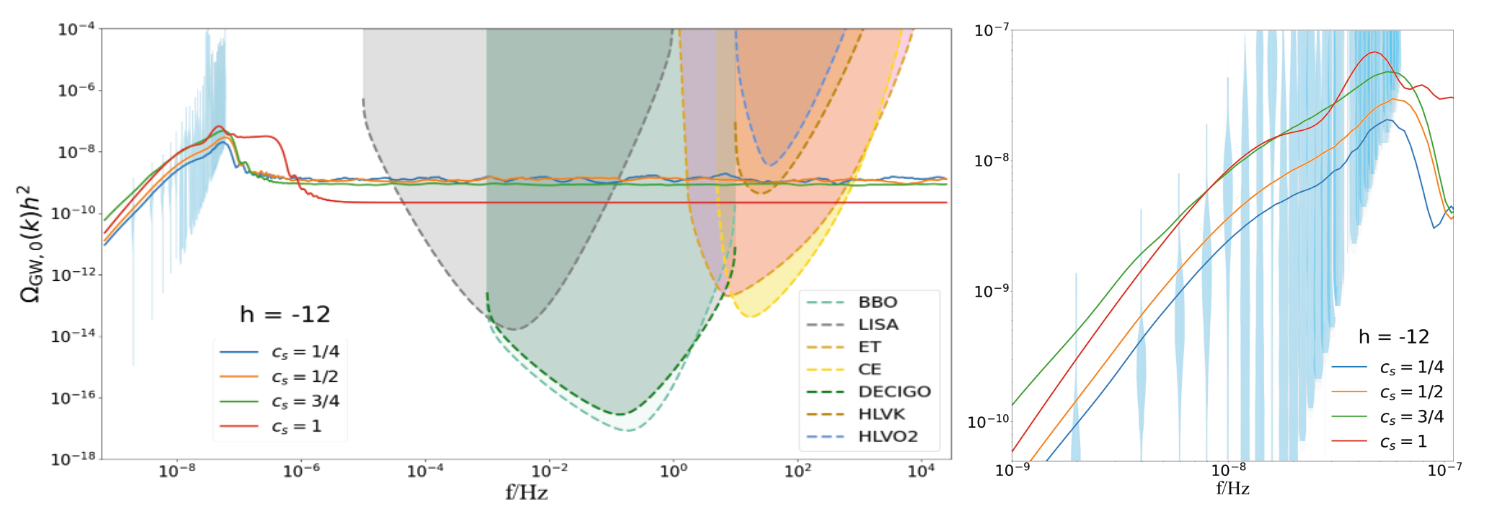}
    \caption{Comparison of the model's results with NANOGrav and future data for $h=-12$. 
    We observe that $c_s=1/4$ has a better fit, particularly with the first violins, compared to other values of sound speed. Furthermore, other values of  $c_s$ can explain the future observations better than  $c_s=1$.}
    \label{Omega-obs-cs-h12}
\end{figure}

The evolution of gravitational wave perturbations follows the same description as presented in \cite{Kohri:2018awv}. Specifically, the evolution of the Fourier mode of gravitational waves can be described as follows
\begin{equation}
\label{GW}
h^{\lambda''}_{k}(\tau)+2\mathcal{H}h^{\lambda'}_{k}(\tau)+k^2h^{\lambda}_{k}(\tau)=4S_k^{\lambda}(\tau),
\end{equation}
Here, $\mathcal{H}=\frac{a'(\tau)}{a(\tau)}$ represents the conformal Hubble parameter, and $S_k^\lambda(\tau)$ corresponds to the source term, which is given by

\begin{equation}
    S_k^\lambda=\int\frac{d^3q}{(2\pi)^3}\epsilon^\lambda_{ij}(\hat k)q^i q^j \bigg(2\Phi_q\Phi_{k-q}+(\mathcal{H}^{-1}\Phi'_q+\Phi_q)(\mathcal{H}^{-1}\Phi'_{k-q}+\Phi_{k-q})\bigg)\,.
\end{equation}
The expression for the source term $S_k^\lambda(\tau)$ involves the polarization tensor $\epsilon_{ij}^\lambda$. Furthermore, the Bardeen potential $\Phi_k$ is related to the curvature perturbation through the following relation
\begin{equation}
    \Phi_k=\frac{2}{3}\Gamma(k\tau)\mathcal{R}_k,
\end{equation}
where $\Gamma(k\tau)$ is the transfer function.  The power spectrum of gravitational waves, corresponding to the solution of equation \eqref{GW}, is given by
\begin{equation}
    \mathcal{P}_h(k,\tau)=2\int_0^\infty dt\int_{-1}^1dsF(s,t)\mathcal{P}_\mathcal{R}(kv)\mathcal{P}_\mathcal{R}(k u),
\end{equation}
where $u=\frac{s+t+1}{2}$ and $v=\frac{t-s+1}{2}$. For details of  $F(s,t)$ see\cite{Kohri:2018awv}. After calculating the power spectrum of gravitational waves, the observed energy density of produced gravitational waves can be obtained as follows
\begin{equation}
    \Omega_{GW}h_0^2= \Omega_{r}h_0^2\bigg(\frac{g_*}{g_{*,e}}\bigg)^{\frac{1}{3}}\Omega_{GW,e}(f),
\end{equation}
with $ \Omega_{r}h_0^2\simeq 4.2\times 10^{-5}$ in which $h_0=H_0/100\text{km}^{-1}$ and $\Omega_r$ is the current radiation energy density.  Also, $f = c k/(2\pi)$ and the factor $\big(\frac{g_*}{g_{*,e}}\big)^{\frac{1}{3}}$ accounts for the change in the number of relativistic degrees of freedom from the end of radiation dominated era to the present time.
In Figure \ref{Omega-obs-cs-h6} and \ref{Omega-obs-cs-h12} , the results of the model are presented in comparison with the NANOGrav and future data. In these figures, we have fixed the sharpness parameter to a specific value but compared the results for different values of $c_s$. 
When comparing different values of $c_s$, it is evident that for $c_s=1/4$ and $c_s=1/2$, the model becomes qualitatively more consistent particularly with the first lines of the NANOGrav data, as they cross the violin diagrams, which have a higher probability according to the analysis in \cite{NANOGrav:2023gor}. Conversely, $c_s=3/4$ lies outside the violin diagrams for the first lines and does not explain the data better than other values of $c_s$. Moreover, as seen in the left panel of Fig. \ref{Omega-obs-cs-h6}, various values of $c_s$ present a slightly different prediction for future data, as $c_s=1/2$ and $c_s=3/4$ lie in a wider range of predictions. However, in Fig. \ref{Omega-obs-cs-h12}, $c_s=1$ lies in a narrower region of future data.

\section{Conclusion}
\label{conclusion}
The study of stochastic gravitational waves has become a valuable tool for investigating the physics of both the early and late universe. In this particular work, we have focused on comparing the predictions of a non-attractor model for gravitational waves with the NANOGrav data from PTAs.

The non-attractor model considered in this study is characterized by three distinct phases of inflation. The first and third phases correspond to slow-roll inflation, which is a well-established paradigm in inflationary cosmology. However, the second phase is unique as it exhibits non-attractor behavior. This non-attractor phase introduces additional dynamics and features into the inflationary process.

By comparing the predictions of the non-attractor model with the NANOGrav data, we aim to assess the model's viability and its ability to explain the observed gravitational wave signals. This analysis provides valuable insights into the physics of the early universe and enhances our understanding of the inflationary dynamics and the generation of primordial gravitational waves. To compare our results with the data from PTAs, we have also taken into account observations of the PBH formation parameter, $f_{\text{PBH}}$. Notably, our findings align well with the constraints imposed by the $f_{\text{PBH}}$ data.

In this study, we have explored a range of values for the sound speed parameter, $c_s$, and the transition parameter, $h$. By considering these various scenarios, we aim to comprehensively investigate the implications of different parameter choices and their impact on the predicted outcomes.
From the plotted results, it is evident that models with $c_s=1/2$ and $c_s=1/4$, exhibit better consistency with the observational data especially in the first lines. However, it is important to note that these comparisons are currently qualitative, and in order to provide a more precise assessment, a quantitative approach will be pursued in future studies. 

Moreover, we have considered the parameter space which have a good prediction for the value of $f_{PBH}$ in the range of NANOGrav frequency. One may consider the cases in which the power spectrum doesn't grow enough. This may make the model more consistent with PTAs data for other values of $c_s$. On the hand one can study  the effects of loop corrections in this study as well and see how loop corrections can affect  the predictions of model for the PTAs data. The other possiblity that can be studied is to consider the case in which the mode functions are not initially in a Bunch-Davis vacuum and see how the predictions change with different choices of vacuum.

To summarize, this study emphasizes that models with lower sound speed parameter values, $c_s$, exhibit more agreement with PTAs data. Nonetheless, a meticulous and comprehensive data analysis is required to achieve a more thorough assessment of the parameter space within the model and make a better fit to the PTAs data. Another aspect worth investigating is the utilization of a broader range of sound speed and sharp transition parameter values to derive more dependable results for comparisons. However, this will require additional computational power. We intend to conduct these analyses in future research endeavors.

In addition to the aforementioned comparisons, we have extended the analytical results presented in \cite{Firouzjahi:2023lzg} to accommodate arbitrary values of the sound speed parameter, $c_s$. Furthermore, we have computed the non-Gaussianity parameter, $f_{\text{NL}}$ in this model as well.
As expected, we find that the non-Gaussianity parameter approaches zero as the transition becomes milder. This aligns with the intuition that a smoother transition would lead to a more Gaussian distribution of primordial fluctuations. By generalizing the analytical results and investigating the non-Gaussianity parameter, we gain a deeper understanding of the statistical properties of the primordial perturbations in our model.

\vspace{0.9cm}
   
 {\bf Acknowledgments:}  We would like to express our deep gratitude to Hassan Firouzjahi for his valuable discussions and meticulous comments, which greatly contributed to the development of this work. We are also thankful to  Alireza Talebian, Mohammad Hossein Namjoo and Sina Hooshangi  for their insightful discussions, which have enriched our understanding of the subject matter. K. A. would like to extend special thanks to the University of SISSA and ICTP for their generous hospitality during the conference "Cosmology in Miramare, 2023," where this paper was in progress.

\end{document}